# The Neuroscientific Basis of Flow: Learning Progress Guides Task Engagement and Cognitive Control

Hairong Lu[1], Dimitri van der Linden[1], Arnold B. Bakker[1,2]

[1] Department of Psychology, Education, and Child studies, Erasmus University Rotterdam, the Netherlands

[2] Department of Industrial Psychology and People Management, University of Johannesburg

Address for correspondence:

Hairong Lu, Department of Psychology, Education, and Child studies, Erasmus University Rotterdam, 3062 PA, Rotterdam, the Netherlands.

Tel: +31 6 3393 0958; Email: lu@essb.eur.nl

**Abstract**

People often strive for deep engagement in activities which is usually associated with feelings of flow: a state of full task absorption accompanied by a sense of control and fulfillment. The intrinsic factors driving such engagement and facilitating subjective feelings of flow remain unclear. Building on computational theories of intrinsic motivation, this study examines how learning progress predicts engagement and directs cognitive control. Results showed that task engagement, indicated by feelings of flow and distractibility, is a function of learning progress. Electroencephalography data further revealed that learning progress is associated with enhanced proactive preparation (e.g., reduced pre-stimulus contingent negativity variance and parietal alpha desynchronization) and improved feedback processing (e.g., increased P3b amplitude and parietal alpha desynchronization). The impact of learning progress on cognitive control is observed at the task-block and goal-episode levels, but not at the trial level. This suggest that learning progress shapes cognitive control over extended periods as progress accumulates. These findings highlight the critical role of learning progress in sustaining engagement and cognitive control in goal-directed behavior.

**Introduction**

Being engaged is critical for achieving goals[1]. Deep engagement, often referred to as state of flow, is a condition of complete focus that is typically linked to optimal performance[2–5]. Previous studies reported that deep engagement states often occur during challenging activities that do not necessarily link to external rewards[4,6,7], like learning, gaming, sports, and creative activities. When engaging in these activities, people tend to perform them for their own sake and care less about external factors. They derive satisfaction and pleasure from the activity itself. Meanwhile, concentration and control under such circumstances are experienced as relatively easy or low-effort [8–10].

Sustaining deep engagement in activities is often challenging in real-life scenarios, where fluctuations in task engagement are common and can hinder goal achievement[11–13]. It is a familiar experience to hear about people struggling to concentrate on their most important objectives, such as students attempting to focus on writing an important paper while being distracted by other activities. Deep engagement necessitates the strategic prioritization of goals, an ability that is known to require cognitive control[14,15]. Cognitive control encompasses the processes that facilitate information processing and adaptive, goal-directed behavior, and it is intimately linked with task engagement[16]. Effective cognitive control allows individuals to persist in goal-directed actions despite distractions[17]. Maintaining cognitive control, however, is usually experienced as effortful that one tends to avoid[18], making it a potential factor relating to fluctuations in task engagement[19]. On the other hand, research suggests that cognitive control and conscious feelings of mental effort are dissociable[20]. Especially, people in a flow state typically experience a high sense of (cognitive) control without feeling too much mental effort[21]. This makes the deep engagement state even more mysterious. Therefore, gaining insight into the intrinsic factors that govern control and foster engagement is important. This understanding could unlock the key to mastering the cognitive mechanisms that underpin deep, sustained engagement, ultimately enhancing productivity and individuals' ability to achieve long-term goals.

Research about intrinsic motivation highlights the importance of learning progress (LP) as an intrinsic motivator in human behavior[22–25]. It suggests that humans track their performance over time and adjust their self-directed explorations to maximize learning progress[26]. This LP-based perspective is supported by human and animal studies. For instance, Ten et al[26] found that, in a free-choice experimental setting, individuals' task selection aligned with an LP-based strategy, indicating that learning progress is strategically used to organize human exploration. Additionally, animal studies show that mesolimbic dopamine, crucial for motivation[27], adjusts in response to learning progress during rats' learning activities[28]. Therefore, learning progress may serve as a fundamental intrinsic motivator that guides exploration and engagement in various activities.

Unlike algorithms that tie intrinsic motivation to reducing prediction errors or uncertainty (which favor high-difficulty tasks)[29,30], LP-based algorithms favor tasks of intermediate difficulty. This preference arises because learning progress tends to be highest at this level, as opposed to tasks that are too easy or too difficult. As a result, people tend to spend more time on tasks that offer greater learning progress, usually those of moderate difficulty. This idea is supported by simulations from Kaplan & Oudeyer[24] and empirical evidence from Ten et al.[26]. As such, the LP-based approach fits with the "optimal challenge" assumption, that is frequently used in research areas such as psychological flow, curiosity, and effort allocation. Studies highlight that aligning task difficulty with personal skill levels is key to achieving deep engagement (i.e., the classical inverted U-shaped function of engagement and difficulty)[31–35]. Building on the link between LP and intermediate difficulty, recent research suggests that learning progress may be the foundation of this optimal challenge, driving engagement[36,37]. In other words, whereas the notion of intermediate task difficulty in engagement mainly refers to characteristics of the task, learning progress has the potential to provide more direct insight into the underlying cognitive and motivational processes. However, to the best of our knowledge, this theory has not been previously tested in a controlled experimental setting, leaving its neurocognitive mechanism less understood.

The current study aims to investigate how learning progress influences engagement and cognitive control allocation in a controlled experimental setting. We hypothesize that individuals are more engaged in goal-directed behaviors in contexts where they experience consistently noticeable progress compared to conditions with low/no progress. This effect can be manifested by levels of subjective flow and low distractibility. Additionally, we propose that individuals strategically allocate cognitive control based on their learning progress.

Efficient task-related preparation, and feedback processing, reflecting cognitive control, are essential for efficient adaptive behavior and goal attainment[38–40]. Task-related preparation, described as a mental suppression of competing information, involves top-down, voluntary maintenance of goals to suppress attention to other stimuli[41]. It has been proven to be an effective predictor of task performance[42]. Additionally, feedback as an important source of information seeking and performance monitoring plays a key role in the bottom-up regulation of adaptive behavior[43,44]. We anticipate that better task-related preparation and enhanced feedback processing will be associated with learning progress.

To examine the effect of learning progress on task engagement, we calculated participants' learning progress in each task session during the experiment and measured subjectively reported flow experience and distractibility after each session, allowing us to examine the relation between learning progress and episodic task engagement. To further test how learning progress is encoded at different stages of adaptive learning processing, we measured EEG during the task, which gives insight into the extent to which learning progress is reflected in signals associated with top-down preparation and bottom-up feedback information processing. Based on previous studies, signals from event-related potentials (ERP) and event-related spectral perturbations (ERSPs) were used as indices of two cognitive processes. First, top-down preparation-related cognitive control was putatively indexed by the contingent negative variation (CNV) occurring prior to the presentation of the target stimulus[45]. More negative CNV indicating enhanced task-related preparation[46,47] was expected to be associated with learning progress. Second, the bottom-up feedback information processing-related cognitive control was putatively indexed by the feedback-locked P3b[48,49]. Larger P3b amplitude reflecting enhanced feedback processing was expected to be associated with learning progress.

Additionally, alpha waves are typically associated with a relaxed, idle state. Studies suggested that an increase in alpha power (alpha synchronization) is associated with the default mode network (DMN) activation, which is believed to support inwardly oriented attention[50] (i.e., self-referencing thinking). On the other hand, a reduction in alpha power (alpha desynchronization) is often interpreted as a sign of increased cortical activity related to attention and information processing[51–54]. Parietal alpha desynchronization is relevant to efficient control in multiple phases of cognitive processing[55–57]. Accordingly, we expect that learning progress modulated cognitive control can be observed by parietal alpha desynchronization in both the preparation[55] and feedback processing[57] phases, denoting a suppressed activation of DMN and enhanced attentional functions. And more importantly, there is a long-lasting discussion about the involvement of DMN in the flow state[32,58,59]. Therefore, the observation of alpha activity may provide additional insights into understanding the neurocognitive mechanism of deep engagement.

**Results**

We analyzed data from 52 participants who performed a "Juice Store" game in a controlled lab setting. In this gamified task, participants were required to serve juice to gnomes by intuitively estimating the amount of juice they desired through a reinforcement learning process (Figure 1a). Each trial began with an empty bottle that is gradually filled with juice. Participants controlled the amount by pressing the space bar to stop the filling. After providing their response, participants received feedback indicating whether their estimate was correct (shown by a juice picture) or incorrect (indicated by a number showing how much higher or lower their estimate was compared to the desired amount).

Building on previous studies[24], learning progress was manipulated in our task design by inducing different prediction error contexts. Prediction error, which generally describes the discrepancy between a predicted outcome and the actual outcome, was operationalized in this study as the absolute distance between the estimate and the desired amount. To create three prediction error (PE) contexts— decreasing PE, constantly low PE, and constantly high PE—three different ordering preferences were assigned to three gnomes (Figure 1b).

The red gnome (context A) changes the requested amount each time it is satisfied. Interacting with the red gnome, participants encountered a new target after each correct estimate, leading to episodes characterized by continuously decreasing prediction errors. An *episode* is defined as the period from the first guess in each block or the first guess after the last correct estimate until the next correct estimate. The blue gnome (context B) consistently orders the same amount of juice. When interacting with the blue gnome, participants experienced episodes with consistently low prediction errors. The yellow gnome (context C) orders a different amount each time, meaning participants could only hit the target by chance. Interactions with the yellow gnome led to episodes with consistently high prediction errors. The simulated prediction error (PE) function across time in an episode for the three contexts is shown in Figure 1c. Accordingly, the LP functions, which were defined as the derivation of the PE function (LP = $f(PE)'$), were calculated and demonstrated in the lower panel in Figure 1c.

To balance task difficulty across three contexts, potentially linked to prediction errors, a strategy was implemented to manage the likelihood of goal attainment at the context level. We assign different target ranges to each context. Based on pilot results, the target range for gnome A was set to 6 pixels, the target range for gnome B was set to 2 pixels, and the target range for gnome

C was set to 14 pixels. Repeated measures ANOVA revealed that this aspect of the design reached the intended goal because the participants' perceived challenge level of the task (Figure 1d, $F(2) = 1.02$, $p = 0.36$, $\eta_p^2 = 0.01$) showed no difference in the three contexts, even though their objective performance in three contexts varies ($F(2) = 95.15$, $p < 0.001$, $\eta_p^2 = 0.51$, decreasing PE context > high PE context > low PE context, all $p$s < 0.001).

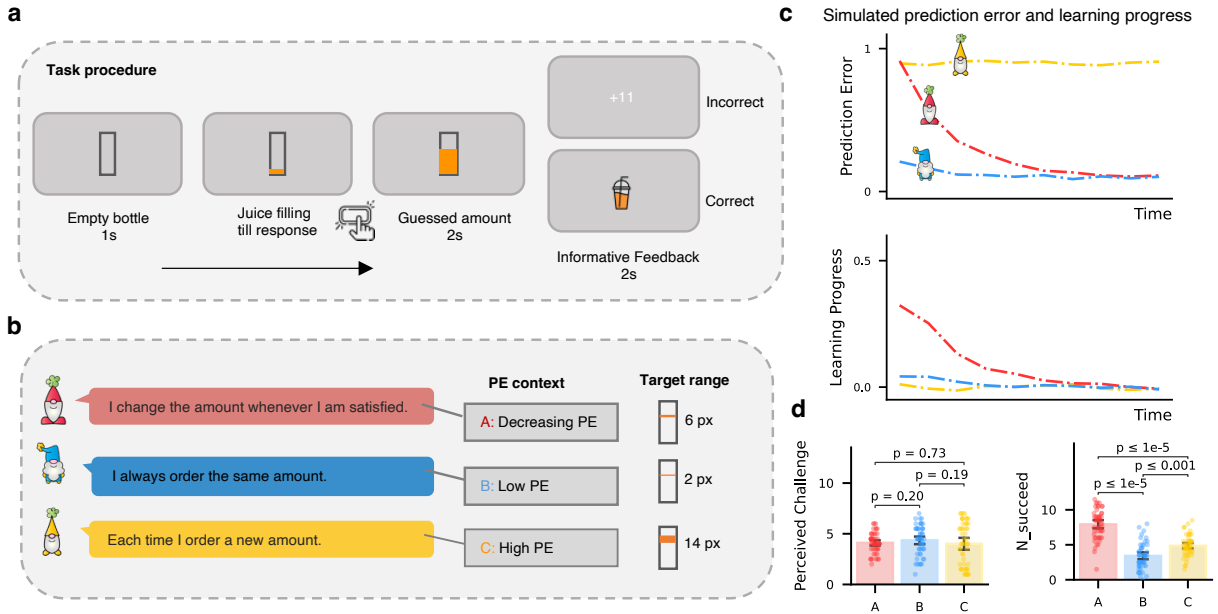

**Figure 1. Experimental design. a,** Participants intuitively estimated the amount of juice requested by gnomes through a reinforcement learning process. Each trial began with an empty bottle displayed on the screen for 1 second, followed by a gradual inflow of juice. Participants pressed the space bar to stop the juice flow at their estimated amount, which remained on the screen for 2 seconds. Feedback was provided afterwards: a juice picture if the estimation was correct, or a number indicating how much higher or lower the estimation was compared to the requested amount. **b,** Participants were randomly assigned to interact with each of the three gnomes twice, completing 30 trials per block. The task was designed to induce three prediction error (PE) contexts: decreasing PE (red gnome, where prediction error decreases over time), constant low PE (blue gnome, consistently low prediction error), and constant high PE (yellow gnome, consistently high prediction error). The gnome character designs were adopted from a previous study[60]. Different target ranges (6 pixels (px) for red gnome, 2 px for blue gnome, and 14 px for yellow gnome) were assigned to three gnomes to counterbalance task difficulty. **c,** Simulated PEs as a function of exploring time (per episode) and the resulting LPs in three experimental conditions. As shown in the lower panel, the reducing PE condition yielded higher learning progress, while consistently high and low PE resulted in lower learning progress in each episode. **d,** Perceived challenge levels across the three PE contexts showed no significant differences, despite participants demonstrating variability in the number of juices served successfully in the game. Context A: decreasing prediction error context; context B: low prediction error

context; context C: high prediction error context. Multiple comparisons were corrected with Benjamin-Hochberg correction.

**Three prediction error contexts yielded different levels of learning progress**

As expected, the interactions with three gnomes generated three PE contexts (A= the decreasing PE context, B= the constant low PE context, and C= the constant high PE context) and yielded high (in context A) and low (in context B and C) LP. As shown in Figure 2a, across time, PE and LP demonstrated a similar pattern as simulated (refer to Figure 1c) The *step* was defined as the sequence of trials to hit a target in our task. Mixed effect models revealed that each step generally reduces PE, but the reduction is more pronounced in context A ($\beta$ = -0.52) compared to contexts B ($\beta$ = -0.14) and C ($\beta$ = -0.09) (see detailed analysis in Supplementary Information). Accordingly, LP (LP = $f(PE)'$) across steps was calculated and demonstrated in the right panel.

Moreover, as shown in Figure 2b, subjective reported perceived progress suggested that participants experienced more progress in a decreasing PE context (A: 5.25 ± 1.22) than a low PE context (3.66 ± 1.38, $t(51)$ = 6.33, $p < 0.001$, $d = 1.22$) and high PE context (2.19 ± 1.18, $t(51)$ = 13.08, $p < 0.001$, $d = 2.54$). Perceived progress in context B was larger than in context C ($t(51)$ = 6.60, $p < 0.001$, $d = 1.14$).

**Learning progress modulates flow experience and subjective-reported distractibility**

We collected subjectively reported flow and distractibility after each task block (see Methods). The findings suggest that higher flow and lower distractibility, indicating a deeper engagement state, happened when interacting with the red gnome (context A) which was associated with greater learning progress.

Repeated measures ANOVA revealed significant main effects of PE contexts on flow experience and distraction (Figure 2c, Flow: $F(2)$ = 75.31, $p < 0.001$, $\eta_p^2$ = 0.372, Figure 2d, distraction: $F(2)$ = 27.30, $p < 0.001$, $\eta_p^2$ = 0.203). Participants experienced significantly higher flow in decreasing PE context (A: 4.69 ± 0.99) than in low PE (B: 4.01 ± 0.82, t(51) = 5.14, p < 0.001, d = 0.75) and high PE (C: 2.99 ± 0.94, $t(51)$ = 10.85, $p < 0.001$, $d = 1.76$) contexts. Flow experience in the low PE context was larger than in the high PE context (t(51) = 7.99, p < 0.001, d = 1.16). A similar effect was found for distractibility, with less distraction in decreasing PE context (A: 3.18 ± 1.24) than in low PE (B: 3.79 ± 1.19, $t(51)$ = -3.18, $p < 0.01$, $d$ = -0.50) and high PE context (C: 4.78 ± 1.50, $t(51)$ = -6.79, $p < 0.001$, $d$ = -1.16). Distractibility in the low PE context was less than high PE context ($t(51)$ = 4.37, $p < 0.001$, $d$ = -0.73).

Results from mixed effect model revealed that learning progress positively predicts flow experience ($b$ = 0.12, SE = 0.017, $\beta$ = 0.47, $p$ < 0.001, Marginal $R^2$ = 0.268, Conditional $R^2$ = 0.392), and negatively predicts distractibility ($b$ = -0.11, SE = 0.023, $\beta$ = -0.34, $p$ < 0.001, Marginal $R^2$ = -0.303, Conditional $R^2$ = 0.337).

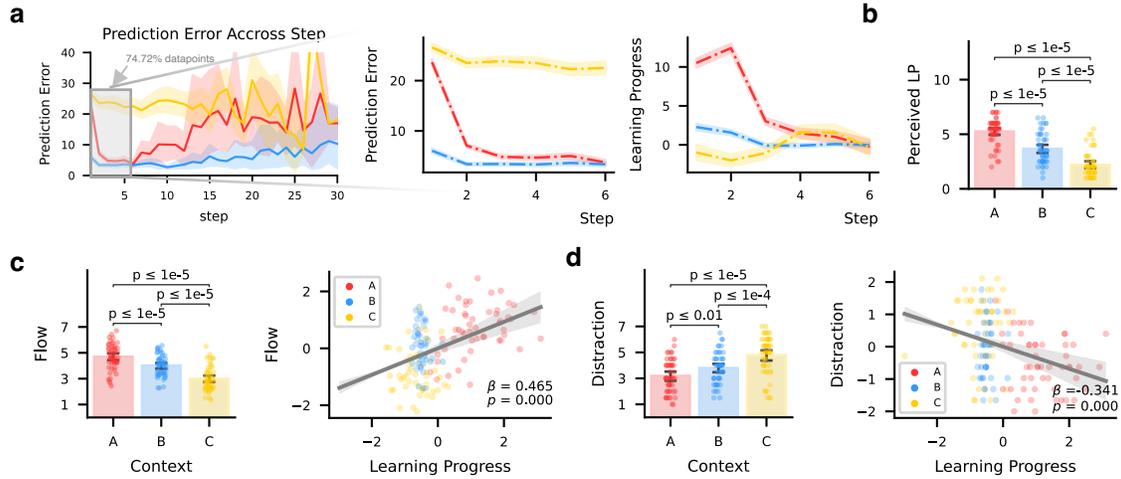

**Figure 2 a,** PE as functions of the step in three experimental conditions. PE was operationalized as the absolute distance between the estimation and the target. The *step* was defined as the sequence of trials to hit a target. We defined the period from the first guess in each block or the first guess after the last correct estimate until the next correct estimate as an *episode*. The average steps needed to hit a target was 5.2 (A=3.4, B=6.9, C=5.3). The first 6 steps (or steps till hit) were extracted from each episode, with 74.72% (6968/9326) data points included in fitting the PE/LP-time models. The second figure shows PE change across the first 6 steps. The third figure shows the calculated LP across the first 6 steps. **b,** Comparisons of perceived LP in three PE contexts. **c,** Flow experience comparisons across PE contexts, and the relation between flow and the calculated objective LP. **d,** Comparisons of subjective reported distractibility in three PE contexts, and the relation between distractibility and the calculated objective LP. **b-d,** Context A: decreasing prediction error context; context B: low prediction error context; context C: high prediction error context. Multiple comparisons were corrected with Benjamin-Hochberg correction.

**Learning progress modulates proactive preparation**

Our behavioral results suggest that learning progress may facilitate deep engagement, because participants experienced more flow and were less distracted when they experienced more learning progress. We also collected EEG data along the task to explore the cognitive features in this deep engagement state that may facilitate goal achievement. In total, 50 participants were included in our cue-locked ERP and ERSPs analysis. We expected the cue-locked CNV and parietal alpha desynchronization, indicating an enhanced preparation, to be associated with LP. CNV was operationalized as the mean amplitude extracted from the time window of 500ms to 1000ms (onset of the juice flow) after cue on FCz electrode[45]. Average parietal alpha power on Pz, P1, and P2 electrodes were extracted from the same time window[51].

Grand averaged cue-locked ERP waveforms in the three PE contexts are shown in Figure 3a. Repeated measures ANOVA for CNV showed a significant main effect of PE contexts (Figure 3b, $F(2) = 3.85$, $p = 0.025$, $\eta_p^2 = 0.04$), where a decreasing PE context (-1.95 ± 1.73) shows a significant larger (more negative) CNV than low (B: -1.37 ± 1.57, $t(49) = -2.57$, $p = 0.025$, $d = -0.35$) and high (C: -1.43 ± 0.87, $t(49) = -2.48$, $p = 0.025$, $d = -0.29$) PE contexts, while there is no difference between the low and high PE context ($t(49) = 0.82$, $p = 0.821$, $d = 0.03$). Results from mixed effect model confirmed a significant negative relation between learning progress and cue-locked CNV amplitude (Figure 3c, $b = -0.049$, SE = 0.022, $\beta = -0.124$, $p = 0.029$, Marginal $R^2 = 0.009$, Conditional $R^2 = 0.684$).

In Figure 3d, we present the cue-related spectral perturbances. Parietal alpha activity extracted from the same time window as CNV revealed the same effects of learning progress. Repeated measures ANOVA for cue-locked parietal alpha shows a significant main effect of PE contexts (Figure 3e, $F(2) = 3.37$, $p = 0.038$, $\eta_p^2 = 0.01$), where a decreasing PE context (A: -0.15 ± 0.15) shows a significant lower parietal alpha oscillation than low (B: -0.12 ± 0.12, $t(49) = -2.33$, $p = 0.036$, $d = -0.28$) and high (C: -0.12 ± 0.15, $t(49) = -2.40$, $p = 0.036$, $d = -0.23$) PE contexts; there is no difference between low and high PE context ($t(49) = 0.18$, $p = 0.85$, $d = 0.02$). Results from mixed effect model revealed a significant negative relation between learning progress and cue-locked parietal alpha activity (Figure 3f, $b = -0.004$, SE = 0.002, $\beta = -0.123$, $p = 0.012$, Marginal $R^2 = 0.995$, Conditional $R^2 = 0.788$).

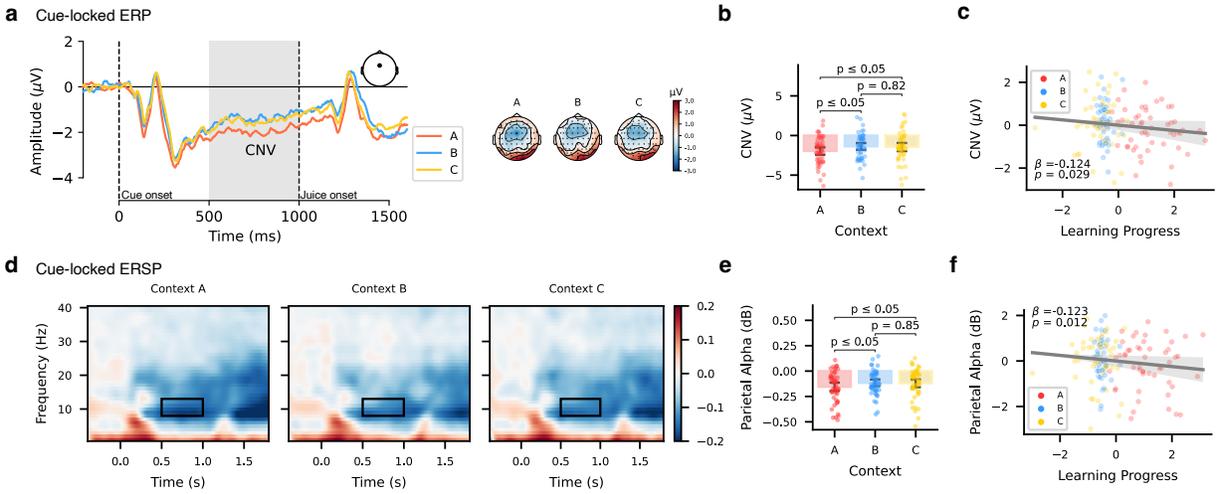

**Figure 3 Learning progress in relation to cue-locked CNV and parietal alpha activity. a,** Grand average ERP waveform locked with the bottle cue on FCz electrode for three PE contexts. As shown in the left topo maps, the negativity during 500s to 1000ms peaked at middle frontal central area. Shaded areas indicate time window used for quantification. **b,** The decreasing PE context show significant lower CNV compared to low and high PE contexts. **c,** Learning progress negatively relates to the cue-locked CNV amplitude. **d,** cue-locked ERSP in parietal region (the average of Pz, P1, P2 electrodes). Marked areas indicate the time-frequency window used for quantification. **e,** Decreasing PE context show significant lower parietal alpha activity compared to low and high PE contexts. **f,** Learning progress negatively relates to cue-locked parietal alpha activity. **a-f,** N=50, Context A: decreasing prediction error context; context B: low prediction error context; context C: high prediction error context. **b and e,** Multiple comparisons were corrected with Benjamin-Hochberg correction.

**Learning progress modulates feedback processing**

EEG data from 50 participants were included in our feedback-locked ERP and ERSP analysis. We expected the feedback-locked P3b and parietal alpha desynchronization, indicating a promoted feedback processing, to be linked with LP. P3b was defined as the mean amplitude extracted from the time window of 400ms to 600ms after the onset of the feedback (only informative feedback (incorrect trials) included). Average alpha power on Pz, P1, and P2 electrodes was extracted from the same time window with P3b.

Grand averaged feedback-locked ERP waveforms in three PE contexts are shown in Figure 4a. Repeated measures ANOVA for P3b shows a significant main effect of PE contexts (Figure 4b, $F(2) = 7.11$, $p = 0.002$, $\eta_p^2 = 0.04$), with decreasing PE context (A: 5.82± 2.32) showing significantly larger (more positive) p3b amplitudes than low (B: 4.90 ± 2.09, $t(49) = 3.68$, $p =$ 0.002, $d = 0.42$) and high (C: 4.80 ± 2.27, $t(49) = 2.91$, $p < 0.001$, $d = 0.44$) PE contexts, while there was no difference between a low and high PE context ($t(49) = 0.33$, $p = 0.740$, $d = 0.04$). Results from mixed effect model confirmed the significant positive relation between learning progress and P3b amplitude (Figure 4c, $b = 0.081$, SE = 0.03, $\beta = 0.16$, $p = 0.006$).

In Figure 4d, we present the feedback-related spectral perturbances. The repeated measures ANOVA for feedback-locked parietal alpha shows a significant main effect of PE contexts (Figure 4e, $F(2) = 3.36$, $p = 0.039$, $\eta_p^2 = 0.02$), with decreasing PE context (A: -0.12 ± 0.16) showing significant lower parietal alpha oscillation than low (B: -0.08 ± 0.14, $t(49) = -2.21$, $p = 0.048$, $d = -0.26$) and high (C: -0.08 ± 0.13, $t(49) = -2.32$, $p = 0.048$, $d = -0.27$) PE contexts, while there is no difference between low and high PE context ($t(49) = -0.06$, $p = 0.951$, $d = 0.01$). Results from mixed effect model revealed a negative relation between learning progress and the feedback-locked parietal alpha (Figure 4f, $b = -0.003$, SE = 0.002, $\beta = -0.100$, $p = 0.057$, Marginal $R^2 = 0.994$, Conditional $R^2 = 0.747$).

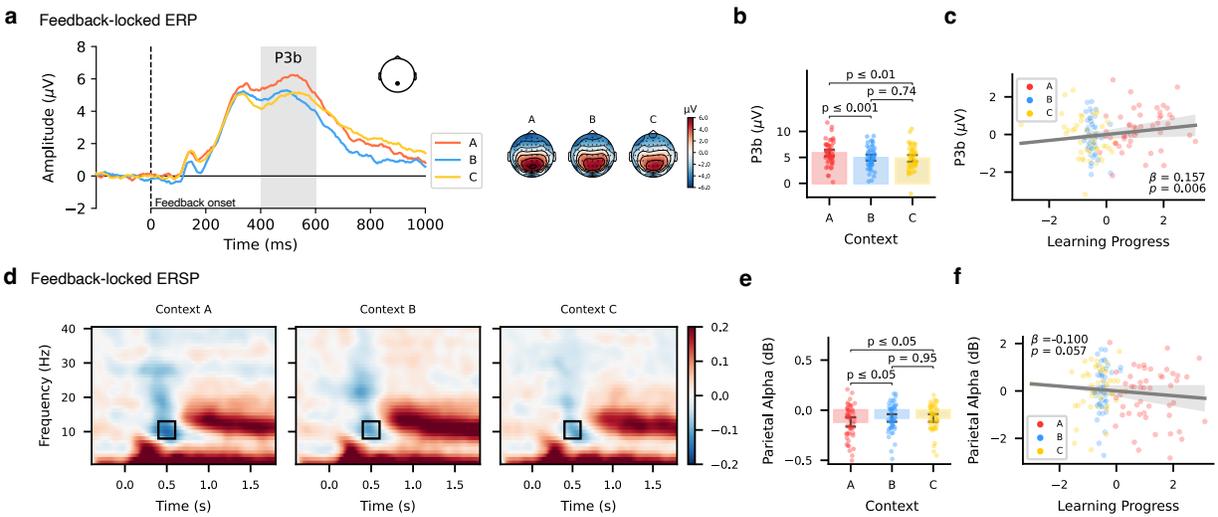

**Figure 4 Learning progress in relation to feedback-locked p3b and parietal alpha activity. a,** Grand averaged ERP waveform locked with the feedback on Pz electrode for three PE contexts. As shown in the topo map, the negativity during 400s to 600ms peaked at middle parietal area. Shaded areas indicate time window used for quantification. **b,** The decreasing PE context show higher P3b amplitude compared with low and high PE contexts. **c,** Learning progress negatively related to the feedback-locked P3b amplitude. **d,** Feedback-locked ERSP on parietal region (the average of Pz, P1, and P2 electrodes). Marked areas indicate the time window-frequency window used for quantification. **e,** Decreasing PE context shows significant lower alpha activity compared to low and high PE contexts. **f,** Learning progress is negatively

related to parietal alpha activity. **a-f,** N=50, Context A: decreasing PE context; context B: low PE context; context C: high PE context. **b and e,** Multiple comparisons were corrected with Benjamin-Hochberg correction.

**Learning progress modulates cognitive control on the task-block and goal-episode level but not necessarily on the trial level**

Our results on the task-block level showed that learning progress regulates task engagement and several aspects of cognitive control. In this section, we conducted analyses on the trial level and goal-episode level respectively to test whether learning progress explains lower-level variability in cognitive control on lower levels. Our analyses were focused on the two ERP components, CNV and P3b. Trial-by-trial learning progress, CNV, and P3b were calculated and fit mixed effect models respectively.

We found that learning progress failed to predict neural signatures of cognitive control on the trial level (Figure 5a, CNV: $b$ = -0.091, $SE$ = 0.090, $\beta$ = -0.013, $p$ = 0.311, Marginal $R^2$ = -36.863, Conditional $R^2$ = 0.066; P3b: $b$ = 0.008, $SE$ = 0.008, $\beta$ = 0.013, $p$ = 0.299, Marginal $R^2$ = -44.91, Conditional $R^2$ = 0.182). However, on the episodic level, we found that learning progress is negatively related to cue-locked CNV and positively related to feedback-locked P3b amplitude (Figure 5b, CNV: $b$ = -0.058, $SE$ = 0.022, $\beta$ = -0.058, $p$ = 0.008, Marginal $R^2$ = 0.382, Conditional $R^2$ = 0.156; P3b: $b$ = 0.058, $SE$ = 0.022, $\beta$ = 0.013, $p$ = 0.008, Marginal $R^2$ = 0.343, Conditional $R^2$ = 0.339). These results indicate that learning progress mainly regulates cognitive control on the higher level (task-block level and goal-episode level) – not on the trial level.

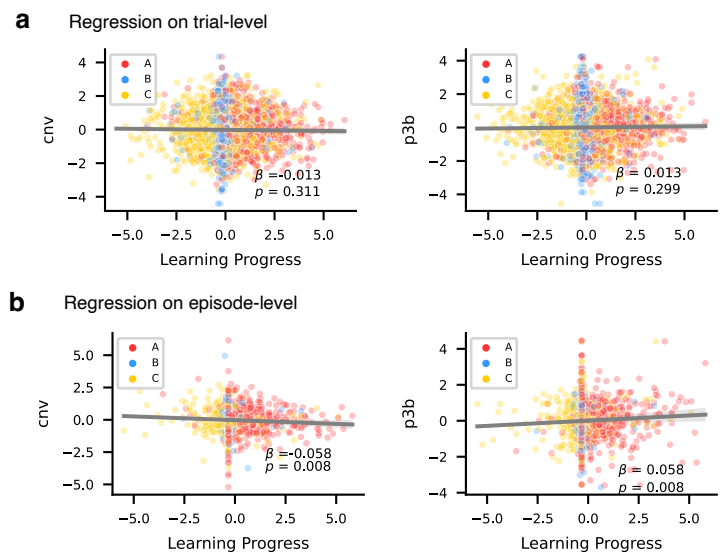

**Figure 5 Learning progress predicting cue-locked CNV and feedback-locked p3b on trial level and episode level. a,** LP failed to predict cue-locked CNV and feedback-locked p3b on trial level. **b,** LP negatively related to cue-locked CNV and positively related to feedback-locked p3b amplitude on episode level.

**Discussion**

Deep engagement, sometimes also referred to as flow, is important for achieving goals efficiently and significantly benefits human well-being[61,62]. Despite the relevance of such levels of engagement, people often struggle to maintain focus and to become fully engaged in important tasks. Building on computational theories of intrinsic motivation, this study focused on learning progress in relation to task engagement and cognitive control in a controlled experimental environment. The findings, supported by behavioral and neural evidence, indicate that learning progress plays a key role in guiding task engagement and cognitive control.

The behavioral data from our study highlight a positive relationship between learning progress and task engagement, as indicated by participants' reported levels of flow and susceptibility to distractions. These results make a case for the notion that deep engagement is likely to occur when people perceive a sufficient level of progress. This insight is relevant because the literature on engagement and flow consistently mentions that flow tends to occur in tasks in which the challenges are matched with the person's skills, such that they are of intermediate difficulty. This presumed inverted U-shape of flow as a function of difficulty was, in fact, replicated by our data (see supplementary information S3)[63,64]. However, the nonmonotonic relation between engagement and difficulty, stating that engagement mainly occurs in tasks with intermediate difficulty does not fully explain the underlying cognitive processes. As we know, human perception of task difficulty is a complex process that involves considering many factors[65–67]. For instance, a task may be perceived as difficult, not only because of the skills required but also because of time pressure, the complexity of instructions, or the level of concentration needed. This makes it challenging to predict the exact circumstance under which someone will experience flow. By addressing a linear relation between learning progress and flow, the current study, on the other hand, explains *why* tasks of intermediate difficulty may yield deeper engagement and provides a cognitive explanation for the emergence of deep engagement.

The EEG findings of our study offer a nuanced understanding of how learning progress relates to cognitive control, particularly through neurophysiological markers associated with top-down proactive control and feedback processing. Specifically, learning progress seems to act as an intrinsic motivator that enhances cognitive control and facilitate the ability to manage and optimize cognitive resources throughout the task performance cycle[14,19,68].

We found that the suppression of pre-stimulus CNV and parietal alpha desynchronization during task preparation were closely tied to the perception of learning progress. CNV is an ERP component observed in the period leading up to an expected stimulus, and it is typically associated with the anticipatory allocation of cognitive resources[69]. In this context, CNV reflects the degree to which an individual is preparing for an upcoming task, engaging in what is known as proactive control[55,70]. The suppression of CNV suggests that individuals who perceive learning progress are indeed more anticipatory, allocating cognitive resources in a manner that aligns with the upcoming task demands before the task even begins. This form of preparation is crucial for optimizing task performance because it ensures that the cognitive system is primed and ready to respond effectively to the relevant stimuli, thereby reducing the cognitive load required during the actual task performance.

The observed parietal alpha desynchronization during task preparation suggests that the Default Mode Network (DMN) is being suppressed, while attention-related functions are being activated. This supports the idea that individuals who feel they are making progress are not just mentally preparing for the task by reducing distractions but are also actively focusing their attention in anticipation of the task[71–73]. This preparatory engagement of attention likely enhances the individual's ability to process task-relevant information more efficiently, leading to better overall performance[55].

The relevance of learning progress is further evidenced by the enhanced feedback-locked P3b amplitude and parietal alpha desynchronization observed during feedback processing. The P3b component is another ERP marker associated with the allocation of attention and the updating of working memory in response to task-relevant stimuli[74]. A larger feedback-locked P3b amplitude indicates a greater allocation of attentional resources to the processing of feedback, which is critical for learning and adaptive behavior[75,76]. This heightened attention to feedback allows for more effective learning because individuals are better able perceive and understand the outcomes of their actions, adjust their strategies accordingly, and refine their approach to achieve their goals[77]. The associated parietal alpha desynchronization during feedback processing further supports this idea, as it indicates that individuals are actively processing the feedback at a deeper level, rather than merely reacting to it[57].

Another novel insight from our study is the observation that the regulatory effects of learning progress on cognitive control manifest primarily at the task block and goal-episode levels, rather than at the trial level. This suggests that intrinsic motivation, driven by perceived progress, operates more effectively over extended periods, where the accumulation of progress is more apparent. It also implies that the cognitive control mechanisms modulated by intrinsic motivation are better suited to managing larger chunks of a task rather than responding to rapid, trial-by-trial fluctuations in task demands. This insight is particularly relevant in real-world settings (i.e., educational and occupational contexts) where tasks are not isolated but form part of a continuous process[21]. This temporal aspect of learning progress aligns with the concept of flow that emerges over prolonged periods of engagement[4]. The fact that flow and enhanced cognitive control are both more likely to occur when individuals perceive progress over time emphasize the inherent interconnectedness of the two processes.

All in all, our findings offer novel insights into the neurocognitive mechanism of deep engagement, which may have implications for neural theories of flow[58,59,78]. For example, one thread of literature suggests that the brain networks responsible for attention and cognitive control (i.e., Central Executive Network (CEN), Salience Network (SN)) are crucial for maintaining focus and concentration during the flow state[6,32,33,58]. Our results align with these ideas, showing that both flow and enhanced cognitive control —evidenced by better pre-stimulus preparation and improved feedback responsiveness— are regulated by learning progress. Additionally, researchers proposed that the Default Mode Network (DMN), typically active during self-referential thinking and mind-wandering, also plays a role in the flow state[32,33,58]. They argue that when individuals are deeply engaged in a task, the DMN becomes less active, leading to a reduction in self-referential thoughts and distraction. We know that alpha oscillations and the DMN are closely connected—when alpha oscillations increase, DMN activity usually strengthens[79]. Our findings show that both the flow state and reduced alpha activity in the parietal region are tied to learning progress. This might mean that during the flow state, the DMN's activity is reduced. Overall, our finding supports a cognitive processing advantage in the flow state.

Future studies may want to expand the present findings by examining how the cognitive and neurological characteristics associated with learning progress may operate in a larger context in relation to deep engagement. For example, though neural characteristics supporting a cognitive processing advantage were found to be related to flow, it is not enough to capture the whole image of flow. The neural mechanism under another important aspect of flow, positive feeling, was left unclear. Future research is expected to explore possible rewarding or 'liking' related neural characteristics in relation to learning progress and flow[58].

Another factor that might be included in future studies is individual differences in tolerance for progress stagnation. Some individuals may maintain engagement and cognitive control even when progress stalls, while others may experience frustration and decreased motivation. Such variation can influence how learning progress affects engagement and cognitive control allocation.

In general, the current study provides empirical evidence that learning progress serves as a powerful intrinsic motivator, driving both engagement and cognitive control. These findings contribute valuable insights into engagement-related theories. While it is often reported that attention and cognitive control are considered effortful and people generally try to avoid effort[18], it is also evident that, in some cases, they tend to seek out and enjoy tasks that require cognitive control[37]. Our study showed that one of the factors facilitating this pleasant cognitive control is experiencing progress.

**Methods**

**Participants**

A total of 54 participants (46 female, 8 male) were recruited from a local university. Participants were between 18 and 26 years of age, with an average age of 20.19 years ($SD$ = 1.89). All participants provided informed consent. All the procedures were approved by the Research Ethics Review Committee of the Department of Psychology, Education & Child Studies, Erasmus University Rotterdam (ETH2324-0444 & ETH2324-0577). Participants were awarded 2.5 credits from participating in this study. We preregistered our sample size, experimental design and analysis plan on https://osf.io/xfkbu

Two participants were excluded from all analyses: one due to reported extreme fatigue, and one who withdrew midway through the experiment. Two participants were excluded further from EEG analysis: one due to loss of EEG data because of maloperation, and one due to bad recording quality (has less than 20% valid trials kept according to our rejection criteria described in EEG preprocessing). The exclusion criteria resulted in 52 participants (44 female, 8 male) in behavioral data analysis and 50 participants (42 female, 8 male) in EEG analysis.

**Procedure**

All participants were instructed to complete a questionnaire, followed by a gamified experimental task with EEG recordings. All tasks were presented on a screen with resolution of 1920 * 1080 pixels and refresh rate of 120 Hz. Participants sit at around 1 meter from the screen. After completing the questionnaire, the EEG recording system was set.

At the beginning of the task, detailed instructions about the experimental task were delivered to participants in an interactive way step by step. First, a cover story of the task was presented saying *'You are now taking the duty to serve juice for three gnomes (red, blue, and yellow). Your challenge is intuitively guessing the quantity of juice they requested. They will leave you a message telling you whether you guessed correctly, or how far away you were from the amount they ordered.'*. Then, participants start with a practice phase. Participants learned how to fill juice in a certain amount (half bottle) and how to interpret the meaning behind the feedback from gnomes in this phase. Participants were told that by pressing the space bar, they could stop the juice flow at any time they wanted. For each trial, participants first saw an empty bottle appearing in the middle of the screen. The bottle was randomly assigned a height ranging between 500 and 550 pixels, and a width ranging between 100 and 200 pixels for each block. The juice then filling in the bottle from the bottom at a speed of 4 pixels per refresh rate (120 Hz). The amount of juice kept growing until the space bar was pressed. The amount of juice participants served was shown on the screen for 2 seconds. Then they saw a number (guess incorrect), or a juice picture (guess correct) on the screen. A positive number means how much more was filled than ordered, while a negative number means how much less was filled than ordered. After three trials of compulsory practice, participants kept practicing until they guessed correctly (see a juice picture).

After the practice phase, participants move on to the free-play phase where they were given 6 opportunities to interact with all gnomes and become familiar with their preferences. Preferences of each gnome were shown on the screen saying: *red gnome: 'I change the amount whenever I am satisfied.'; blue gnome: 'I always order the same amount.'; yellow gnome: 'Each time I order a new amount.'*. By pressing 1, 2, or 3, participants could decide which gnome they want to interact with at each round. Then, same trial procedures as the practice phase were presented. Each round of interaction in the free-play phase consists of 15 trials. After completing each round, a celebration screen with the number of bottles of juice served successfully, and a firework cartoon picture was shown.

After the free-play phase, participants started a duty phase, where they were assigned to randomly serve different gnomes for 30 trials each for two rounds. Trial procedures are the same as the practice phase and free-play phase. There were in total 6 rounds of duty serving. The order was semi-random (random (red, blue, yellow), random (red, blue, yellow)).

In the duty phase, after each round of serving, participants were asked to report their feeling during the round. Questions about levels of flow experience, distraction, experienced progress, and the amount of challenge felt were delivered on a 7-point Likert scale. After reporting their feeling, a celebration screen with the number of bottles of juice served successfully, and a firework cartoon picture was shown.

**EEG recording and preprocessing**

EEG data were recorded from the 64 BioSemi Active-electrodes (Ag-AgCI) electrodes embedded in a stretched Lycra cap (with the 10/20 international system layout) with a sampling rate of 512 Hz. The Active-electrode is a sensor with a very low output impedance, all problems with regards to capacitive coupling between the cable and sources of interference, as well as any artifacts by cable and connector movements are completely eliminated (https://www.biosemi.com/active_electrode.htm). Vertical electrooculography (VEOG) was recorded from two electrodes placed above and below the right eye. Horizontal electrooculography (HEOG) was recorded from two electrodes placed lateral to the external canthi at both sides. As per BioSemi's design, the ground electrode during acquisition was formed by the Common Mode Sense active electrode and the Driven Right Leg passive electrode. Signals were amplified using the BioSemi ActiveTwo system.

EEG data processing was conducted using the MNE-Python package[80]. During preprocessing, continuous data were notch filtered at 50 Hz and band-pass filtered at 0.1 Hz and 100 Hz. EEG data further high pass filtered at 1 Hz for the purpose of conducting independent component analysis (ICA)[81]. The 1 Hz high pass filtered data were decomposed into independent components using the infomax independent component analysis (ICA) algorithm implemented in MNE. Independent components were inspected fully automatically using the function from the MNE-ICALabel package. Components that were not classified as 'brain' or 'other' were excluded. Basically, all components excluded were labeled 'eye', 'muscle', or 'channel noise'. ICA weights obtained from the 1 Hz high pass filtered data were then applied back to the original data. EEG data were re-referenced to the average of all electrodes.

Pre-processed EEG data were epoched relative to the onset of cue (empty bottle) and feedback events: cue-locked epochs (-200 to 2000 ms), feedback-locked epochs (-200 to 1000 ms). All epochs were baseline-corrected using the mean amplitude before event onset (-200 to 0 ms). Epochs containing artifacts with amplitudes exceeding ± 100 μV were excluded from further analysis[82]. We focused our analyses on these event related potentials, ROIs and time windows determined a priori based on the literature[83].

The EEG spectral power was assessed by calculating the event-related spectral perturbations (ERSPs). Time-frequency representations (TFRs) of power were estimated on the EEG epochs using Morlet wavelets[84]. We analyzed frequencies ranging from 1 to 40 Hz and these frequencies were linearly spaced to provide a comprehensive analysis across the spectrum. The number of cycles for each frequency was set to be half the frequency value (e.g., 2 cycles for 4 Hz, 10 cycles for 20 Hz). This approach balances temporal and spectral resolution[85]. Fast Fourier Transform (FFT) was employed to speed up the computation process. A decimation factor of 3 was applied to reduce the computational load by down sampling the data. This means every third data point was retained, effectively reducing the temporal resolution while preserving the overall signal characteristics. The baseline period was defined from -0.5 to -0.2 seconds relative to the event onset[86]. The correction mode used was 'logratio', which computes the logarithm of the ratio between the power during the baseline period and the power at each time point. This method helps in normalizing the power values and highlighting relative changes.

**Variables and Analysis**

*Prediction error (PE)* was calculated to measure the accuracy of participants' predictions. PE was defined as the discrepancy between the predicted value ($p_{t-1}$) and the actual outcome ($r_{t-1}$) in previous trial [87]. It was illustrated by the equation: $PE_t = |r_{t-1} - p_{t-1}|$. It was operationalized as the discrepancy in the percentage of the bottle between the guessing amount and the requested amount in the current study. Here, $r_{t-1}$ is the requested amount and $p_{t-1}$ is the participant's guessed amount from the previous trial. This measure captures the magnitude of the discrepancy, providing a clear metric for assessing prediction accuracy.

*Learning progress (LP)* was defined as the derivative of the prediction error function: $LP = f(PE)$ [87]. Mean LP was calculated for each condition as the behavorial probe of progress manipulation check. Besides, we also measured perceived progress and challenge level as psychological probes of the progress manipulation check. Perceived Progress was measured using one item 'I had the feeling that I made progress'. Challenge level was measured using on item 'How challenging did you feel?'. The items were delivered on a 7-point Likert scale (1=not at all, 7= very much).

*Flow experience* was measured using 7 items adapted from previous studies [88,89]. The items are '1) I felt I was in the zone and didn't have to force myself to concentrate; 2) I love the feeling of what I was doing, and want to capture this feeling again; 3) My thought and actions seemed to happen automatically; 4) Time passed quickly in this round; 5) I knew clearly what to do at each step; 6) I had a sense of control; 7) I felt the right amount of challenge. To prevent the potential triggering of self-referential related mind-wandering, we excluded a self-referential thinking item from the original scale in our measurement [90]. All items were delivered on a 7-point Likert scale (1=not at all, 7= very much). Cronbach alpha of the flow experience scale is 0.88. Mean score of the seven items were calculated as the index of flow experience.

*Distractibility* was measured as an additive opposite index of task engagement using one item adapted from Austin and Hemsley [91]: 'I easily got distracted'. It was delivered on a 7-point Likert scale (1=not at all, 7= very much).

*Task performance* was indexed by the number of juices served successfully in each condition.

*CNV* was extracted from the time window of 500-1000ms after the onset of the empty bottle (which is -500-0 ms before the onset of the juice flow) on FCz electrode[45]. The CNV amplitude was calculted as the mean amplitude of the interested time window.

*P3b* was extracted from the time window of 400-600 ms after the onset of informative feedback (incorrect trials only) on Pz electrode[48]. The P300 amplitude was calculated as the mean amplitude of the interested time window.

*Parietal alpha* during multiple task execution phases were calculated as the mean power of the Pz, P1, and P2 electrodes in frequency window of 8-13 Hz[51].

The *step* was defined as the sequence of trials to hit a target in our task. An *episode* in our task was defined as the period from the first guess in each block or the first guess after the last correct estimate until the next correct estimate. The average steps needed to hit a target was 5.2 (A=3.4, B=6.9, C=5.3). The first 6 steps (or steps till hit) were extracted from each episode, with 74.72% (6968/9326) data points included in fitting the PE/LP-time models.

Repeated measures ANOVAs were conducted to test the effect of PE contexts on interested variables. Post hoc comparisons were corrected with Benjamini-Hochberg correction[92]. Mixed effect models[93] with subject number entered as a random effect were conducted to test the relation between learning progress and interested variables. The Q-Q plots were inspected for any deviations from normally distributed residuals, and we confirmed that no significant collinearity existed between the regressors.

10. Ullén, F., de Manzano, Ö., Theorell, T. & Harmat, L. The Physiology of Effortless Attention: Correlates of State Flow and Flow Proneness. in *Effortless Attention* 205–218 (The MIT Press, 2013). doi:10.7551/mitpress/9780262013840.003.0011.

11. Esterman, M., Noonan, S. K., Rosenberg, M. & DeGutis, J. In the Zone or Zoning Out? Tracking Behavioral and Neural Fluctuations During Sustained Attention. *Cerebral Cortex* **23**, 2712–2723 (2013).

12. Hopstaken, J. F., Van der Linden, D., Bakker, A. B. & Kompier, M. A. J. The window of my eyes: Task disengagement and mental fatigue covary with pupil dynamics. *Biological Psychology* **110**, 100–106 (2015).

13. Kane, G. A. *et al.* Increased locus coeruleus tonic activity causes disengagement from a patch-foraging task. *Cognitive, Affective and Behavioral Neuroscience* **17**, 1073–1083 (2017).

14. Bugg, J. M. & Egner, T. The many faces of learning-guided cognitive control. *Journal of Experimental Psychology: Learning, Memory, and Cognition* **47**, 1547–1549 (2021).

15. Eisma, J., Rawls, E., Long, S., Mach, R. & Lamm, C. Frontal midline theta differentiates separate cognitive control strategies while still generalizing the need for cognitive control. *Sci Rep* **11**, 14641 (2021).

16. Botvinick, M. M. & Cohen, J. D. The Computational and Neural Basis of Cognitive Control: Charted Territory and New Frontiers. *Cognitive Science* **38**, 1249–1285 (2014).

17. Marsh, J. E., Sörqvist, P. & Hughes, R. W. Dynamic cognitive control of irrelevant sound: Increased task engagement attenuates semantic auditory distraction. *Journal of Experimental Psychology: Human Perception and Performance* **41**, 1462–1474 (2015).
25